\documentstyle{mn}
\input{epsf}

\title[Dark matter in dwarf spheroidals]{Dark matter distribution in
dwarf spheroidal galaxies}

\author[Ewa L. {\L}okas]{Ewa L. {\L}okas\\ Nicolaus Copernicus Astronomical
Center, Bartycka 18, 00--716 Warsaw, Poland\\ E-mail: lokas@camk.edu.pl}

\begin{document}

\maketitle

\begin{abstract}
We study the distribution of dark matter in dwarf spheroidal galaxies by
modelling the moments of their line-of-sight velocity distributions. We discuss
different dark matter density profiles, both cuspy and possessing flat
density cores. The predictions are made in the framework of standard
dynamical theory of two-component (stars and dark matter) spherical systems
with different velocity distributions. We compare the predicted velocity
dispersion profiles to observations in the case of Fornax and Draco dwarfs.
For isotropic models the dark haloes with cores are found to fit the data
better than those with cusps. Anisotropic models are studied by fitting two
parameters, dark mass and velocity anisotropy, to the data. In this case all
profiles yield good fits but the steeper the cusp of the profile, the more
tangential is the velocity distribution required to fit the data. To resolve this
well-known degeneracy of density profile versus velocity anisotropy we obtain
predictions for the kurtosis of the line-of-sight velocity distribution for
models found to provide best fits to the velocity dispersion profiles. It turns
out that profiles with cores typically yield higher values of kurtosis which
decrease more steeply with distance than the cuspy profiles, which will allow to
discriminate between the profiles once the kurtosis measurements become available.
We also show that with present quality of the data the alternative explanation of
velocity dispersions in terms of Modified Newtonian Dynamics cannot yet be ruled
out.
\end{abstract}


\section{Introduction}

Dark matter is now generally believed to be the dominating component of the matter
content of the Universe
which has played an important role in the formation of structure. Significant
effort has gone into establishing its properties and mechanisms of evolution. The
theoretical results which got most attention in the recent years concerned the
shape of density profiles of dark matter haloes. Those were mainly obtained by the
means of $N$-body simulations with the most popular being the so-called universal
profile advocated by Navarro, Frenk and White (1997, hereafter NFW). NFW found (and
others confirmed) that in a large range of masses the density profiles of
cold dark matter haloes forming in various cosmologies can be fitted
with a simple formula with only one fitting parameter. This density profile
steepens from $r^{-1}$ near the centre of the halo to $r^{-3}$ at large
distances. More recent simulations with higher resolution produce even steeper
density profiles, with inner slopes $r^{-3/2}$  (Fukushige \& Makino 1997;
Moore et al. 1998; see also Jing \& Suto 2000).

Analytical calculations also predict density profiles with steep inner cusps.
The NFW profile is claimed to be reproduced in studies taking into account
the merging mechanism (Lacey \& Cole 1993) in the halo formation scenario (NFW,
Avila-Reese, Firmani \& Hernandez 1998). On the other hand, using an extension
of the spherical infall model \L okas (2000) and \L okas
\& Hoffman (2000) have shown that if only radial motions are allowed the inner
profiles have to be steeper than $r^{-2}$ (in agreement with considerations
based on distribution functions, see Richstone \& Tremaine 1984). Although addition
of angular momentum is known to make profiles shallower (White \& Zaritsky 1992;
Hiotelis 2002), its origin is not yet clear enough to make exact predictions.

The expected steep inner profiles are now being tested by comparison with
observations of density profiles of galaxies and
galaxy clusters. Some recent studies of clusters (e.g. van der Marel et al. 2000)
claim good agreement between cluster observations and the NFW mass density
profile, while other based on lensing find the dark matter to exhibit a smooth
core (Tyson, Kochanski \& dell'Antonio 1998, see also Williams, Navarro \&
Bartelmann 1999). In the case of galaxies, however, the situation is even
more troubling. Flores \& Primack (1994) and Moore (1994) were the first
to indicate that the NFW profile is incompatible with the rotation curves of
spiral galaxies. More and more evidence is being presented now, which shows that
spiral galaxies possess a dark matter core rather than a cusp in the inner parts
(Burkert 1995; Salucci \& Burkert 2000; Borriello \& Salucci 2001).
The evidence is especially convincing in the case of LSB galaxies which are supposed
to be dominated by dark matter (de Blok et al. 2001; de Blok, McGaugh \& Rubin 2001).
Some mechanisms to explain the formation of such core have also been proposed
(Hannestad 1999; Kaplinghat, Knox \& Turner 2000; El-Zant, Shlosman \& Hoffman 2001;
Ravindranath, Ho \& Filippenko 2001). However, as emphasized by van den Bosch
\& Swaters (2001) even good quality rotation curves cannot always discriminate
between constant density cores and cusps because of the effects of beam smearing,
the uncertainties in the stellar mass-to-light ratio and the limited spatial
sampling of the halo density distribution.

Dwarf spheroidal (dSph) galaxies provide a unique testing tool
for the presence and distribution of dark matter because due to their large velocity
dispersions they are believed to be dominated by this component (for a review
see Mateo 1997). Measurements of central velocity dispersion allowed many authors
to estimate mass-to-light ratios in dwarfs and find it to be much larger than
stellar values (see Mateo 1998a and references therein). These calculations were made
with the simplifying assumption of isotropic velocity distribution and in most cases
of mass following light. With the
measured velocity dispersion profiles becoming available now, this assumption can
in fact be relaxed and much more information can be extracted from the data. However,
as in all pressure supported systems, when trying to model the velocity dispersion
dependence on distance we are faced with the well-known degeneracy of density
profile versus velocity anisotropy (Binney \& Tremaine 1987).

The purpose of this work is to study moments of the line-of-sight velocity
distribution following from different dark matter density profiles 
and different assumptions concerning velocity anisotropy. We focus on modelling
the velocity dispersion profiles and extend here the earlier work (\L okas 2001),
which discussed only NFW profile, by considering both cuspy profiles and those
possessing flat cores, in order to decide which of them (and under what conditions)
are favoured by observational data in the case of dSph galaxies. In order to 
discriminate between different models we also obtain predictions for the fourth-order
velocity moment. Our work is complementary to the approach of Wilkinson et al. (2002) 
and Kleyna et al. (2001, 2002) who discussed dark matter distributions differing at
large distances from the centre.

The paper is organized as follows. In Section~2 we briefly present the method of
calculation of the line-of-sight velocity moments in models with different
density profiles and velocity anisotropy. Section~3 describes the assumptions about
the matter content of dSph galaxies. Results for the velocity dispersion profiles
of Fornax and Draco dwarfs for both isotropic and anisotropic velocity distribution
are given in Section~4, where we also present predictions for the fourth moment,
the kurtosis. Section~5 provides comments on the possible alternative
explanation of velocity dispersions in terms of Modified Newtonian Dynamics (MOND)
and the discussion follows in Section~6.

\vspace{-0.3in}

\section{Moments of the velocity distribution}

The dynamics of stars in a gravitational potential $\Phi(r)$ is completely
described by the distribution function $f({\bf r}, {\bf v})$ governed by the
Boltzmann equation. It is often advantageous however to work not with $f$ but
with its velocity moments defined as
\begin{equation}    \label{d6}
    \nu \overline{v_r^i v_\theta^j v_\phi^k}
    = \int v_r^i v_\theta^j v_\phi^k f(r, {\bf v}) {\rm d}^3 v ,
\end{equation}
where $\nu(r)$ is a 3D density of stars. In the following we will assume that
the system is spherically symmetric and that there are no net streaming motions
(e.g. no rotation). Then all the odd velocity moments vanish.

At second order the two distinct moments
are $\overline{v_r^2}$ and $\overline{v_\theta^2}=\overline{v_\phi^2}$ which we will
denote hereafter by $\sigma_{\rm r}^2$ and $\sigma_\theta^2$ respectively.
They are related by the lowest order Jeans equation
derived from the Boltzmann equation (Binney \& Tremaine 1987)
\begin{equation}    \label{m1}
        \frac{\rm d}{{\rm d} r}  (\nu \sigma_{\rm r}^2) + \frac{2 \beta}{r} \nu
	\sigma_{\rm r}^2 + \nu \frac{{\rm d} \Phi}{{\rm d} r} =0,
\end{equation}
where
\begin{equation}    \label{d7}
	\beta=1-\frac{\sigma_\theta^2(r)}{\sigma_{\rm r}^2(r)}
\end{equation}
is a measure of the anisotropy in the velocity distribution and $\Phi$ is the
gravitational potential.

The most convenient approach to solving equation (\ref{m1}) for
$\sigma_{\rm r}^2$ is to make an assumption about $\beta$. The values
of this parameter have been studied both observationally and via
$N$-body simulations but only in systems very different from dSph galaxies.
For dark matter haloes Thomas et al. (1998) find that, in a variety of
cosmological models, the ratio $\sigma_\theta/\sigma_{\rm r}$ is not far
from unity and decreases slowly with distance from the centre to reach
$\simeq 0.8$ at the virial radius. Observations of elliptical galaxies are
also consistent with the orbits being isotropic near the centre and
somewhat radially anisotropic farther away, although cases with tangential
anisotropy are also observed (Gerhard et al. 2001). In our considerations
the anisotropy parameter describes the properties of the observed motion of
stars which are only a tracer population of the underlying dark matter
potential and therefore we cannot assume their anisotropy is similar to that
of dark haloes or elliptical galaxies. In the absence of theory of their formation,
which could provide some hints, or direct measurements of velocity anisotropy
in dSph galaxies we have to consider different values of $\beta$.

Traditionally, $\beta$ has been modelled in a way
proposed by Osipkov (1979) and Merritt (1985) where
$\beta_{\rm OM} = r^2/(r^2 + r_{\rm a}^2)$,
with $r_{\rm a}$ being the anisotropy radius determining the transition
from isotropic orbits inside to radial orbits outside. This model covers
a wide range of possibilities from isotropy ($r_{\rm a} \rightarrow
\infty, \beta=0$) to radial orbits ($r_{\rm a} \rightarrow 0, \beta=1$).
Another possibility
is that of tangential anisotropy with the fiducial case of circular orbits
when $\beta \rightarrow - \infty$. For our purposes here it is most convenient to
cover all possibilities from radial to isotropic and circular orbits with
a simple model of $\beta=$const and $-\infty < \beta \le 1$.

The solution of the Jeans equation (\ref{m1}) with the boundary condition
$\sigma_{\rm r} \rightarrow 0$ at $r \rightarrow \infty$ for $\beta$=const is
\begin{equation}	\label{m4b}
	\nu \sigma_{\rm r}^2 (\beta={\rm const})= r^{-2 \beta}
	\int_r^\infty r^{2 \beta} \nu \frac{{\rm d} \Phi}{{\rm d} r} \ {\rm d}r .
\end{equation}
From the observational point of view, an interesting, measurable quantity
is the line-of-sight velocity dispersion obtained from the 3D velocity
dispersion by integrating along the line of sight (Binney \& Mamon 1982)
\begin{equation}    \label{m3}
    \sigma_{\rm los}^2 (R) = \frac{2}{I(R)} \int_{R}^{\infty}
    \left( 1-\beta \frac{R^2}{r^2} \right) \frac{\nu \,
    \sigma_{\rm r}^2 \,r}{\sqrt{r^2 - R^2}} \,{\rm d} r \ ,
\end{equation}
where $I(R)$ is the surface brightness.

Introducing result (\ref{m4b}) into equation (\ref{m3})
and inverting the order of integration the calculations of $\sigma_{\rm
los}$ can be reduced to one-dimensional numerical integration of a formula
involving special functions for arbitrary $\beta=$const.

It has been established that by studying $\sigma_{\rm los} (R)$ alone
we cannot uniquely determine the properties of a stellar system. In fact
systems with different densities and velocity anisotropies can produce
identical $\sigma_{\rm los} (R)$ profiles (see e.g. Merrifield \& Kent 1990;
Merritt 1987). As will be shown below, the same kind of degeneracy also
plagues the modelling of dSph galaxies. It is therefore interesting to
consider higher-order moments of the velocity distribution hoping that the
models with the same $\sigma_{\rm los} (R)$ will differ in their projected
higher-order moments.

In the case of fourth-order moments we have three distinct components
$\overline{v_r^4}$, $\overline{v_\theta^4}=\overline{v_\phi^4}$ and
$\overline{v_r^2 v_\theta^2}=\overline{v_r^2 v_\phi^2}$ related by two
higher order Jeans equations (Merrifield \& Kent 1990). In order to calculate
the moments we need additional information about the distribution function.
According to Jeans theorem the anisotropic distribution functions can be
constructed as functions of the integrals of motion, the energy $E=-\Phi-v^2/2$
and angular momentum $L=(v_\theta^2 + v_\phi^2)^{1/2} r$. In the isotropic case
the distribution function is a function of energy alone and can be determined
uniquely from the underlying mass distribution.
However, as discussed by Dejonghe (1987), in the anisotropic case,
in which we will be most interested here, the function is not uniquely determined
by the assumed mass density, even if information on velocity dispersions is
available.

Therefore we will restrict ourselves here to functions which can be constructed
from the energy-dependent distribution function by multiplying it by some function
of angular momentum. We will adopt the form
\begin{equation}    \label{d8}
	f(E, L) = f_0 (E) L^{-2 \beta}
\end{equation}
with $\beta=$const, a generalization first considered by H\'enon (1973).
The velocity moments of such a function can be explored without explicit calculation
of $f_0 (E)$. For example, using equation (\ref{d6}) one can show
that the function of the form (\ref{d8})
has the desired property of $\sigma_\theta^2/\sigma_{\rm r}^2 = 1-\beta$.
Similar calculation proves that the three fourth-order moments are related by
\begin{eqnarray}
	\overline{v_r^2 v_\theta^2} & = & \frac{1}{3} (1-\beta) \overline{v_r^4}
	\label{d9} \\
	\overline{v_\theta^4} & = & \frac{1}{2} (1-\beta) (2-\beta) \overline{v_r^4}
	\label{d10} .
\end{eqnarray}
Then the two Jeans equations for the fourth-order moments reduce to one of the form
\begin{equation}    \label{d11}
        \frac{\rm d}{{\rm d} r}  (\nu \overline{v_r^4}) + \frac{2 \beta}{r} \nu
	\overline{v_r^4} + 3 \nu \sigma_{\rm r}^2 \frac{{\rm d} \Phi}{{\rm d} r} =0 ,
\end{equation}
similar to equation (\ref{m1}). The solution, in analogy to equation (\ref{m4b}), is
\begin{equation}	\label{d12}
	\nu \overline{v_r^4} (\beta={\rm const})= 3 r^{-2 \beta}
	\int_r^\infty r^{2 \beta} \nu \sigma_{\rm r}^2 (r)
	\frac{{\rm d} \Phi}{{\rm d} r} \ {\rm d} r .
\end{equation}

By projection we obtain the line-of-sight fourth moment
\begin{eqnarray}
    \overline{v_{\rm los}^4} (R) &=& \frac{2}{I(R)} \int_{R}^{\infty}
    \left[ \left( 1- \frac{R^2}{r^2} \right)^2 + 2(1-\beta)
    \right. \label{d13} \\
    &\hspace{-2.8cm} \times & \hspace{-1.7cm} \left. \frac{R^2(r^2-R^2)}{r^4}
    + \frac{(2-\beta)(1-\beta)}{2} \frac{R^4}{r^4} \right]
    \frac{\nu \,  \overline{v_r^4} \,r}{\sqrt{r^2 - R^2}} \,{\rm d} r \ ,
    \nonumber
\end{eqnarray}
where we used relations (\ref{d9})-(\ref{d10}). Introducing equations (\ref{d12})
and (\ref{m4b}) into (\ref{d13}) and inverting the order of integration the
calculation of $\overline{v_{\rm los}^4} (R)$ can be reduced to two-dimensional
numerical integration. A useful way to express the fourth projected moment is to
scale it with $\sigma_{\rm los}^4$ in order to obtain the kurtosis
\begin{equation}	\label{d14}
	\kappa_{\rm los} (R) = \frac{\overline{v_{\rm los}^4} (R)}
	{\sigma_{\rm los}^4 (R)}
\end{equation}
which will be discussed in Subsection 4.3.

\section{Matter content}

We will now discuss different mass distributions contributed by stars
and dark matter to the gravitational potential $\Phi$ in equation (\ref{m1}).

\subsection{Stars}

The distribution of stars is modelled in the same way as in \L okas (2001)
i.e. by the S\'ersic profile (S\'ersic 1968, see also Ciotti 1991)
\begin{equation}    \label{m5}
	I(R) = I_0 \exp [-(R/R_{\rm S})^{1/m}],
\end{equation}
where $I_0$ is the central surface brightness and $R_{\rm S}$ is the
characteristic projected radius of the S\'ersic profile. The best-fitting S\'ersic
parameter $m$ has been found to vary in the range $1 \le m \le 10$
(Caon, Capaccioli \& D'Onofrio 1993) for different elliptical galaxies, however
for dSph systems $m=1$ is usually used, although in some
cases other values of $m$ are found to provide better fits
(e.g. Caldwell 1999).

The 3D luminosity density $\nu(r)$ is obtained from $I(R)$ by
deprojection
\begin{equation}	\label{m5a}
	\nu(r) = - \frac{1}{\pi} \int_r^\infty \frac{{\rm d} I}{{\rm d} R}
	\frac{{\rm d} R}{\sqrt{R^2 - r^2}}.
\end{equation}
In the case of $m=1$ we get $\nu(r, m=1) = I_0 K_0 (r/R_{\rm S})/(\pi R_{\rm S})$,
where $K_0(x)$ is the modified Bessel function of the second kind. For
other values of $m$ in the range $1/2 \le m \le 10$ an excellent
approximation for $\nu(r)$ is provided by (Lima Neto, Gerbal \&
M\'arquez 1999)
\begin{eqnarray}
	\nu(r) &=& \nu_0 \left( \frac{r}{R_{\rm S}} \right)^{-p}
	\exp \left[-\left( \frac{r}{R_{\rm S}}
	\right)^{1/m} \right] \label{m5c} \\
	\nu_0 &=& \frac{I_0 \Gamma(2 m)}{2 R_{\rm S}
	\Gamma[(3-p) m]}  \nonumber \\
	p &=& 1.0 - 0.6097/m + 0.05463/m^2 \nonumber.
\end{eqnarray}

The mass distribution of stars following from (\ref{m5c}) is
\begin{equation}	\label{m5d}
	M_{*}(r) = \Upsilon L_{\rm tot} \frac{\gamma[(3-p)m,
	(r/R_{\rm S})^{1/m}]}{\Gamma[(3-p)m]},
\end{equation}
where $\Upsilon$=const is the mass-to-light ratio for stars, $L_{\rm tot}$
is the total luminosity of the galaxy and $\gamma(\alpha, x) = \int_0^x
{\rm e}^{-t} t^{\alpha-1} {\rm d} t$ is the incomplete gamma function.

\subsection{Dark matter}

We explore here different density distributions of dark matter which can
be described by the following general formula
\begin{equation}    \label{m6}
    \rho(r) = \frac{\rho_{\rm char}}{(r/r_{\rm
    s})^\alpha \,(1+r/r_{\rm s})^{3-\alpha}} ,
\end{equation}
where $\rho_{\rm char}$ is a constant characteristic density. As is immediately
clear from equation (\ref{m6}), the inner slope of the profile is $r^{-\alpha}$
while the outer one is always $r^{-3}$.
We choose the parameter $\alpha$ in the range $0 \le \alpha \le 3/2$, which covers
a wide range of possible inner profiles. The cuspy profiles of $\alpha>0$ are
motivated by the results of $N$-body simulations. The profile with $\alpha=1$
corresponds to the so-called universal profile proposed by NFW (see also
\L okas \& Mamon 2001) as a fit to the profiles of simulated haloes, while the
profile with $\alpha=3/2$ is identical to the one following from higher resolution
simulations of Moore et al. (1998). The profile with $\alpha=0$,
possessing an inner core is favoured by observations of spiral galaxies
(see the Introduction) and is very similar (but not identical) to the profile
proposed by Burkert (1995).

The scale radius $r_{\rm s}$ introduced in equation (\ref{m6}) marks the distance
from the centre of the object
where the slope of the profile is the average of the inner and outer slope:
$r^{-(3+\alpha)/2}$. Additional parameter which controls the shape of the profile
is the concentration which we define here in a traditional way as
\begin{equation}    \label{d1}
    c=\frac{r_v}{r_{\rm s}},
\end{equation}
where $r_v$ is the virial radius, i.e. the distance from the centre
of the halo within which the mean density is $v = 200$ times the
present critical density, $\rho_{\rm crit,0}$. The density of $200 \rho_{\rm crit,0}$
is traditionally assumed by most $N$-body simulators to be the mean density of the
object collapsing and virializing at the present epoch. This value, following from the
spherical collapse model, depends however on the underlying cosmology, e.g. $v=178$ is
valid for the Einstein-de Sitter model, while we have $v \approx 100-120$
for the currently
most popular $\Lambda$(Q)CDM models with $\Omega_M=0.3$ and $\Omega_{\Lambda(Q)}=0.7$
(\L okas \& Hoffman 2001a, 2001b). However, since this study is motivated by the
dark halo profiles obtained in $N$-body simulations we keep $v=200$ in order to have
the same definition of concentration parameter as in most simulations.

We normalize the density profile (\ref{m6}) so that the mass within $r_v$
is equal to the so-called virial mass
\begin{equation}    \label{d2}
    M_v=\frac{4}{3} \pi r_v^3 v \rho_{\rm crit,0}.
\end{equation}
The characteristic density of equation (\ref{m6}) then becomes
\begin{equation}    \label{d3}
    \rho_{\rm char}=\frac{(3-\alpha) v \rho_{\rm crit,0} \ c^\alpha}{3 F(c)},
\end{equation}
where $F(c)$ is given by the hypergeometric function
\begin{equation}    \label{d4}
    F(x)= \ _2 F_1 (3-\alpha, 3-\alpha; 4-\alpha; -x).
\end{equation}
The dark mass distribution following from (\ref{m6}), (\ref{d2}) and (\ref{d3}) is
\begin{equation}    \label{d5}
    M_{\rm D}(s) = M_v s^{3-\alpha} \frac{F(c s)}{F(c)},
\end{equation}
where we introduced $s=r/r_v$.

The concentration of simulated dark matter haloes has been observed to
depend on the virial mass. Jing \& Suto (2000) tested the relation $c (M_v)$
for the masses of the
order of normal galaxies and clusters in the case of density profiles with
$\alpha=1$ and $\alpha=3/2$ and found the concentration to decrease with mass
with the highest $c \approx 10$ ($\alpha=1$) and $c \approx 5$ ($\alpha=3/2$)
observed for the mass scale of the order of a normal galaxy
($10^{12} h^{-1} M_{\sun}$).
In the previous work (\L okas 2001) we used an extrapolation of the $c (M_v)$
relation fitted to the simulations results to smaller masses characteristic
of dwarf galaxies. Here, however, similar relation would be needed for the $\alpha=0$
profile and the information provided by observations is very scarce. Although
relations e.g. between the inner mass and core radius have been found in the case
of Burkert profile (Burkert 1995) and those can be translated into an $c (M_v)$
relation useful here, the estimates of this kind have been performed only for a
small number of spiral galaxies and the errors of the fitted parameters are large
(Borriello \& Salucci 2001). We have therefore decided to assume different
concentration parameters $c=$ const in the further analysis. If we believe, however,
that the trend of $c$ decreasing with $M_v$ extends to smaller masses (and the relation
was found to be even steeper for subhalos by Bullock et al. 2001) we should expect
the cuspy halos in dSph galaxies to have $c > 10$ ($\alpha=1$) and $c > 5$
($\alpha=3/2$).

\section{Results for the Fornax and Draco dwarfs}

In the two following subsections we model the velocity dispersion profile of Fornax
and Draco first assuming isotropy and then for arbitrary anisotropy
parameter $\beta$=const.
In the third subsection we present predictions for the line-of-sight kurtosis for some
of the best-fitting models.

\subsection{Isotropic orbits}

{\samepage
\begin{figure}
\begin{center}
    \leavevmode
    \epsfxsize=7.8cm
    \epsfbox[50 50 320 560]{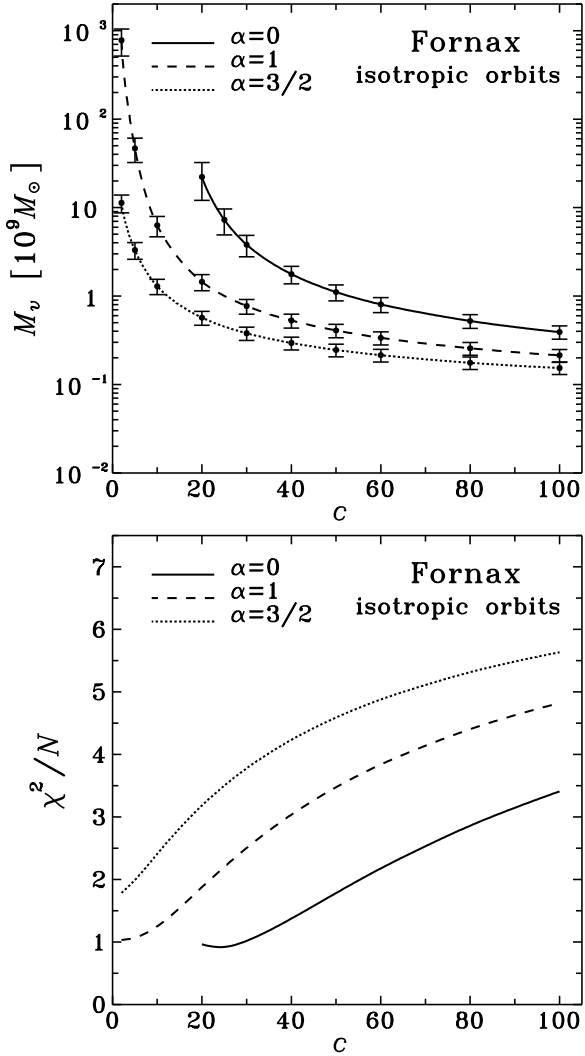}
\end{center}
\caption{Upper panel: the best fitting masses of dark haloes as a function of
assumed concentration parameter for profiles with different $\alpha$ in the case
of isotropic orbits for the Fornax dwarf. Lower panel: the goodness of fit
measure, $\chi^2/N$, for the fits in the upper panel.}
\label{mvchiif}
\end{figure}

\begin{figure}
\begin{center}
    \leavevmode
    \epsfxsize=7.8cm
    \epsfbox[50 50 320 560]{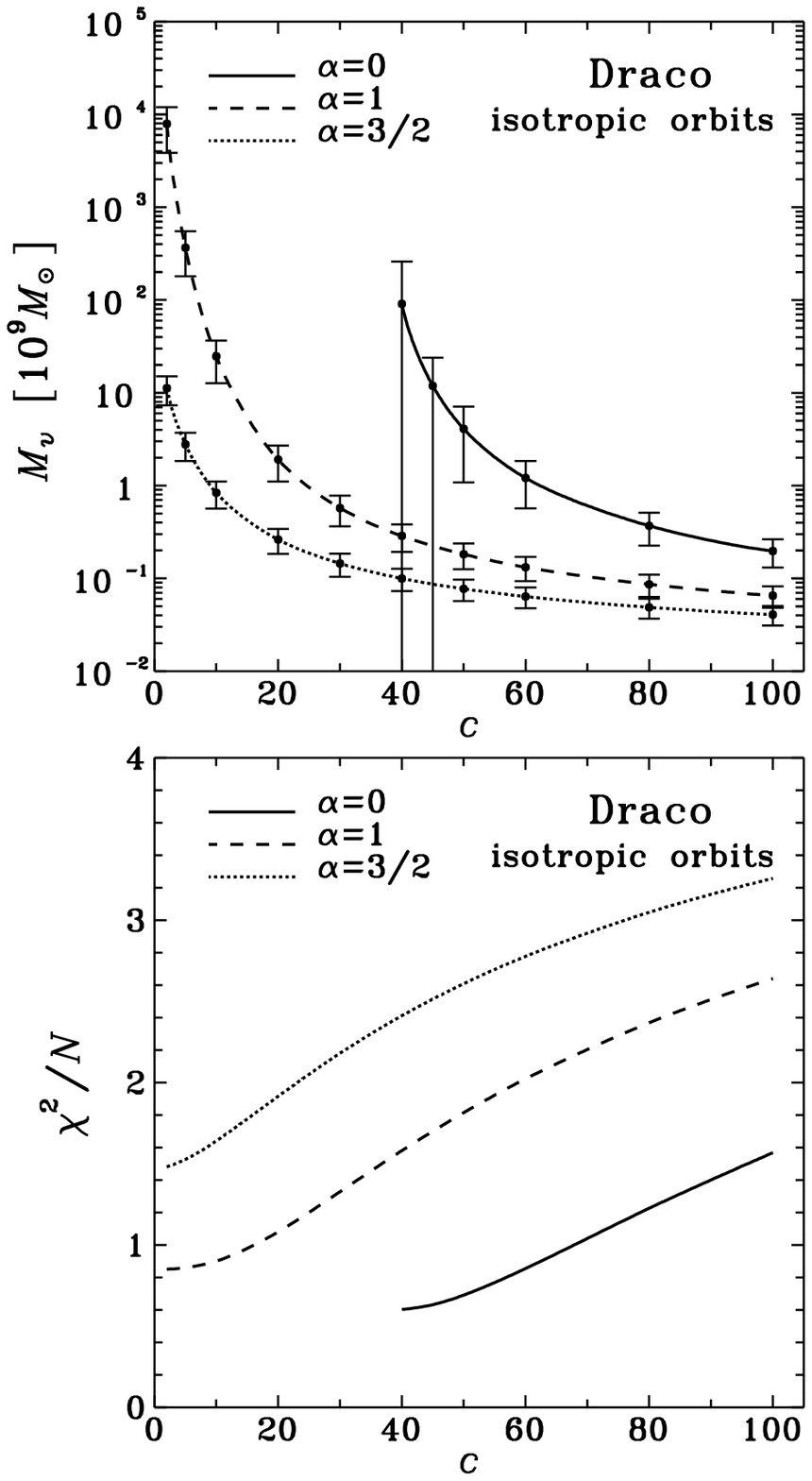}
\end{center}
\caption{Same as Figure~\ref{mvchiif}, but for Draco.}
\label{mvchiid}
\end{figure}
}

{\samepage
\begin{figure}
\begin{center}
    \leavevmode
    \epsfxsize=7.8cm
    \epsfbox[50 50 320 560]{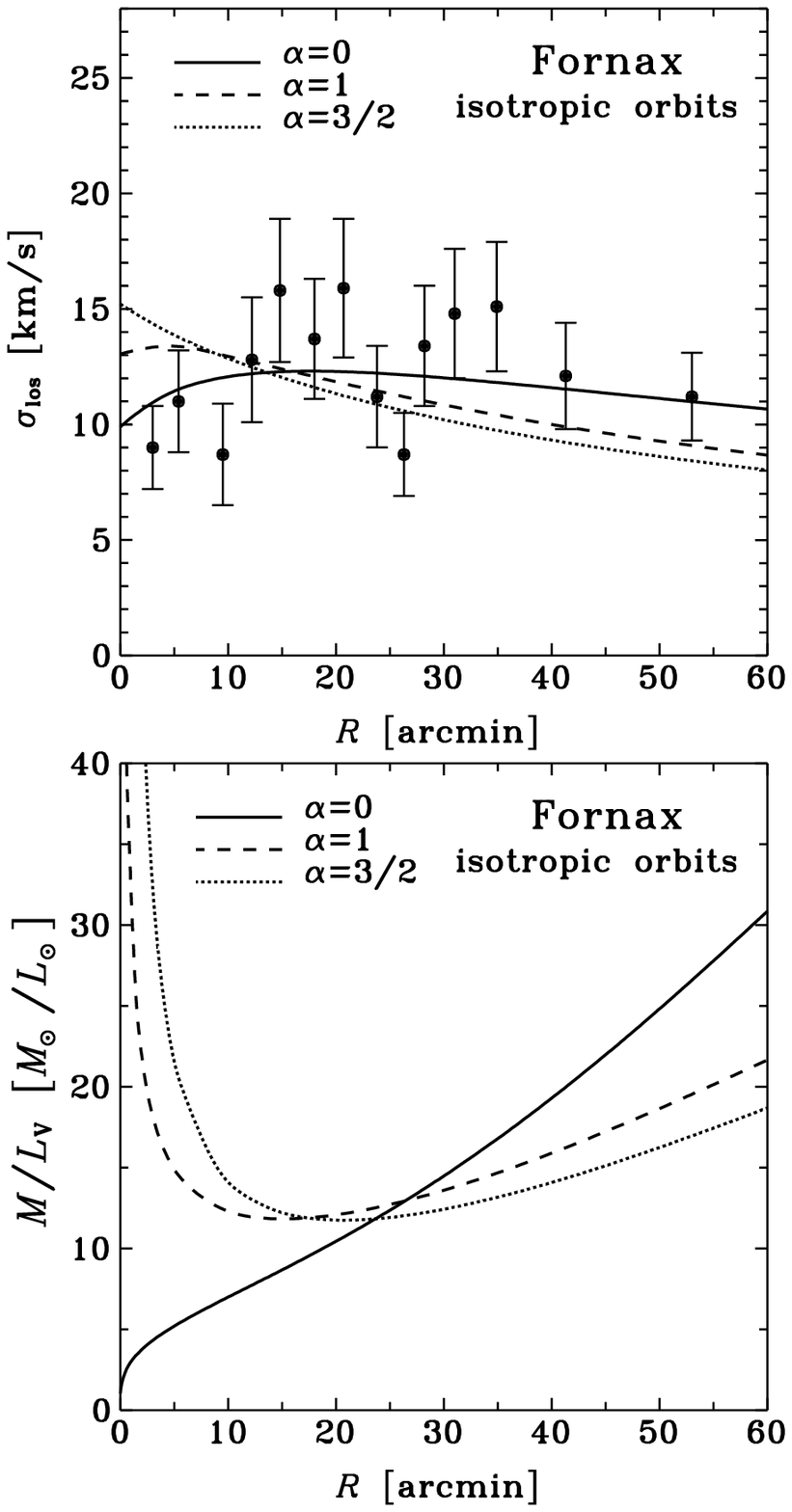}
\end{center}
\caption{Upper panel: the best fitting velocity dispersion profiles
obtained for $c=30$ ($\alpha=0$), $c=20$ ($\alpha=1$) and $c=10$ ($\alpha=3/2$)
in the case of isotropic orbits for the Fornax dwarf. Lower panel:
mass-to-light ratios for the fits in the upper panel.}
\label{fisomtl}
\end{figure}

\begin{figure}
\begin{center}
    \leavevmode
    \epsfxsize=7.8cm
    \epsfbox[50 50 320 560]{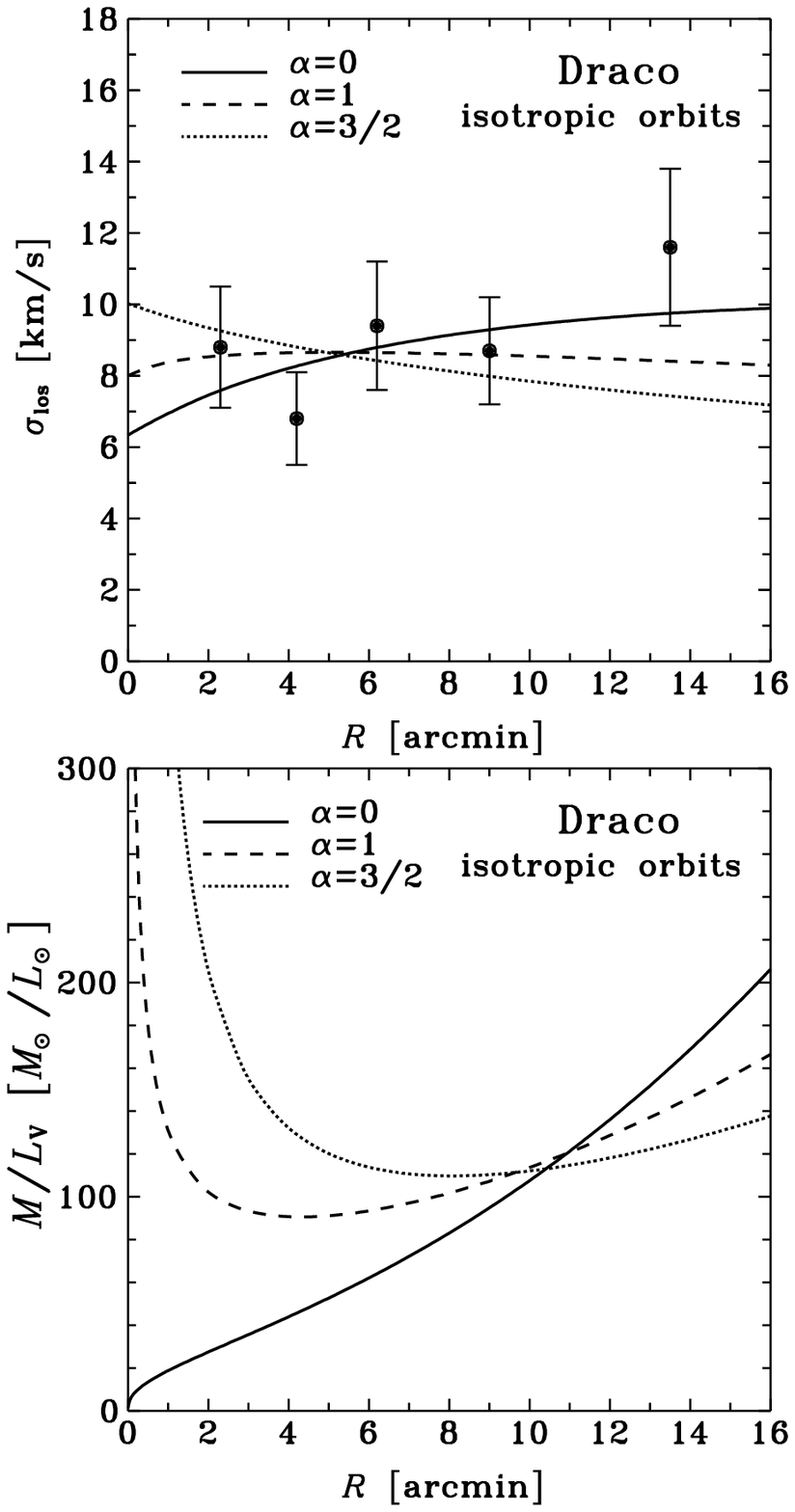}
\end{center}
\caption{Same as Figure~\ref{fisomtl}, but for Draco. The best fitting
velocity dispersion profiles plotted were obtained for $c=50$ ($\alpha=0$),
$c=20$ ($\alpha=1$) and $c=10$ ($\alpha=3/2$). }
\label{disomtl}
\end{figure}
}

The observational parameters of the Fornax and Draco dwarfs needed in the following
computation were taken from Irwin \& Hatzidimitriou (1995) and are
summarized in Table~\ref{tab1}. The Table gives the fitted S\'ersic radius
$R_{\rm S}$ assuming the $m=1$ S\'ersic distribution of equation (\ref{m5}),
the distance modulus $m-M$, the distance in kpc with its error,
the brightness of the galaxies
in V-band, $M_{\rm V}$, and the total luminosity in V-band, $L_{\rm tot,V}$,
together with its error. We adopt the mass-to-light ratio
for stars in this band to be $\Upsilon_{\rm V} \approx 1 M_{\sun}/L_{\sun}$
(Mateo et al. 1991).

\begin{table}
\caption{Observational parameters of the Fornax and Draco dwarfs
from Irwin \& Hatzidimitriou (1995).}
\label{tab1}
\begin{center}
\begin{tabular}{ccc}
parameter & Fornax & Draco \\
\hline
$R_{\rm S}$(arcmin)  & 9.9   & 4.5  \\
$R_{\rm S}$(kpc)     & 0.35  & 0.094  \\
$m-M$                & 20.4  & 19.3 \\
distance (kpc)       & $120 \pm 8$  & $72 \pm 3$ \\
$M_{\rm V}$(mag)     & -13.0 & -8.3  \\
$L_{\rm tot,V}$($L_{\sun}$)  & $(1.4 \pm 0.4) \times 10^7$ &
$(1.8 \pm 0.8) \times 10^5$  \\
\hline
\end{tabular}
\end{center}
\end{table}

In the previous work (\L okas 2001) it was shown that the density of stars
alone cannot produce the observed velocity dispersion profiles in the case
of Fornax and Draco. In the following we will explore the hypothesis
that the velocity dispersions are generated by stars moving in the Newtonian
gravitational acceleration generated by stars and dark matter.

We first assume that the velocity distribution of stars is isotropic, i.e. we
consider models obtained from equations (\ref{m4b}) and (\ref{m3}) in the
case of $\beta=0$. The 3D distribution of stars $\nu(r)$ is given by
(\ref{m5c}). The gravitational acceleration is
$g=- {\rm d} \Phi/{\rm d} r= -G M(r)/r^2$ with
$M(r)=M_*(r) + M_{\rm D}(r)$, where $M_*(r)$ and $M_{\rm D}(r)$
are given by equations (\ref{m5d}) and (\ref{d5}), respectively. Using such
modelling of the velocity dispersion profile
we perform a one-parameter fitting of the velocity dispersion data.
In the case of Fornax we use the data of Mateo (1997) and in the case of Draco
those of Armandroff, Pryor \& Olszewski 1997 (based on the observations of
Armandroff, Olszewski \& Pryor 1995). In each case we assume the value of the
concentration parameter $c$ and fit by the method of least squares (Brandt 1999)
the dark virial mass $M_v$ of the models
with different dark matter profiles distinguished by the parameter $\alpha$ of
equations (\ref{m6}) and (\ref{d5}).

The resulting best-fitting values of $M_v$ as a function of assumed $c$ are shown
in the upper panels of Figures~\ref{mvchiif} and \ref{mvchiid}. The error bars
indicate 1$\sigma$ errors in $M_v$ associated with the errors in the velocity
dispersion measurements. They are obtained from the covariance matrix
(in this case one-element) of the fitted parameters.
The errors in mass in the whole range of
concentrations shown in Figures~\ref{mvchiif} and \ref{mvchiid} are of the
order of 16-45\% for Fornax and 24-185\% for Draco. The lower panels of these
Figures plot $\chi^2/N$, the measure of the goodness of fits presented in the
upper panels, where $N$ is the number of degrees of freedom (in this case we
have $N=13$ for Fornax and $N=4$ for Draco). The results
for $\alpha=0$ do not reach to very low concentration because the least-square
fitting converges more and more slowly there yielding incredibly high dark masses.
One might think that a two-parameter $(c, M_v)$ fitting would be a better choice
here, but as the lower panels of Figures~\ref{mvchiif} and \ref{mvchiid} show, this
would give the lowest possible $c$ in all cases. In the case of cuspy profiles
resulting from $N$-body simulations low $c$ of the order of unity is never
obtained except for objects of the mass of cluster of galaxies. Clearly for both
Fornax and Draco the $\alpha=0$ profile provides the best fit in terms of
$\chi^2/N$. Although quite low $\chi^2/N$ can be also obtained for $\alpha=1$ profile,
this happens only for unrealistic low concentrations and yielding incredibly high
mass of a halo of the order of a normal galaxy.

The predictions for the line-of-sight velocity dispersion profiles of the
best-fitting models with $\alpha=0, 1$ and $3/2$ in the
isotropic case are shown in the upper panels of Figures~\ref{fisomtl} and
\ref{disomtl} for Fornax and Draco respectively together with the data.
For clarity of the illustration we have chosen only one of the fits from
Figures~\ref{mvchiif} and \ref{mvchiid} for each type of profile. In the case of
$\alpha=0$ profile no a priori information on $c$ is available so we have chosen
the value giving the best fit and reasonably low best-fitting mass $M_v$ at the
same time, i.e. $c=30$ in the case of Fornax and $c=50$ in
the case of Draco. For cuspy profiles the concentrations expected on the basis
of the $N$-body results were chosen (see Subsection 3.2).
We have taken $c=20$ for $\alpha=1$ and $c=10$ for
$\alpha=3/2$. The best-fitting virial masses obtained for these concentrations
in the isotropic case together with 1$\sigma$ error bars following from the
uncertainties in velocity dispersion measurements are given in the
upper part of Table~\ref{parameters}. The method of least squares applied here
allows also to study the propagation of errors associated with other observables entering
the models. Including the errors in luminosity and distance given in Table~\ref{tab1}
and assuming an error in the stellar mass-to-light ratio of $0.5 M_{\sun}/L_{\sun}$ (an
estimate given by Mateo et al. (1991) in the case of Fornax) we obtain errors for
$M_v$ which are given in brackets in Table~\ref{parameters}.

\begin{table}
\caption{Examples of best-fitting parameters in the isotropic and
anisotropic case with
1$\sigma$ error bars. $M_v$ is given in units of $10^9 M_{\sun}$.}
\label{parameters}
\begin{center}
\begin{tabular}{ccccc}
case & $\alpha$ & parameter & Fornax & Draco \\
\hline
$\beta=0$ &  0  & $M_v$   & $3.8 \pm 1.0(1.2)$  & $4.1  \pm 3.0(3.2) $ \\
          &  1  & $M_v$   & $1.5 \pm 0.3(0.3)$  & $1.9  \pm 0.8(0.8) $ \\
          & 3/2 & $M_v$   & $1.3 \pm 0.3(0.3)$  & $0.84 \pm 0.27(0.27) $ \\
\hline
$\beta \ne 0$ & 0 & $M_v$   & $ 3.4 \pm 1.0(1.1)$ & $ 4.3  \pm 3.4(3.6) $ \\
              &   & $\beta$ & $-0.36 \pm 0.33$ & $-0.11 \pm 0.47 $ \\
              & 1 & $M_v$   & $ 1.3 \pm 0.3(0.3)$ & $ 2.4  \pm 1.1(1.2) $ \\
              &   & $\beta$ & $-1.5 \pm 0.8$ & $-0.98  \pm 1.2 $ \\
              &3/2& $M_v$   & $ 1.2 \pm 0.3(0.3)$ & $ 1.2  \pm 0.4(0.4) $ \\
              &   & $\beta$ & $-2.6 \pm 1.6$ & $-3.5  \pm 5.1 $ \\
\hline
\end{tabular}
\end{center}
\end{table}

Lower panels of Figures~\ref{fisomtl} and \ref{disomtl} show the combined
mass-to-light ratio in V-band for Fornax and Draco
calculated from
\begin{equation}	\label{m12}
	M/L_{\rm V} = \frac{M_{\rm D}(r) + M_* (r)}{L_{\rm V}(r)},
\end{equation}
where $M_{\rm D}(r)$ and $M_* (r)$ are given
by equations (\ref{d5}) and (\ref{m5d})
respectively, and $L_{\rm V}(r) = M_* (r)/\Upsilon_{\rm V}$
is the luminosity distribution. Calculation of $M/L_{\rm V}$ were made for the
same models as presented in the upper panels of the Figures.
The growth of $M/L_{\rm V}$ at large distances is due to the behaviour
of $M_{\rm D} (r)$ which diverges logarithmically for all considered $\alpha$
while the mass in stars is finite. However, while the cuspy profiles produce
divergent $M/L_{\rm V}$ at the centre of the galaxy, the $\alpha=0$ profile
gives a finite $M/L_{\rm V}$ increasing at all radii.

All the mass-to-light estimates shown
give a similar $M/L_{\rm V} \approx 10 M_{\sun}/L_{\sun}$
in the case of Fornax and $M/L_{\rm V} \approx 100 M_{\sun}/L_{\sun}$ in the case
of Draco at a scale of the order of two S\'ersic radii and are consistent with
the values of the central mass-to-light ratios given by Mateo et al. (1991) and
Armandroff et al. (1997).

Kleyna et al. (2001) estimated mass-to-light ratio for
Draco using their new velocity measurements (Kleyna et al. 2002). Assuming isotropic
orbits and the velocity
dispersion to be constant or a linear function of distance they calculated the mass
needed to produce such velocities from the Jeans equation. The resulting $M/L_{\rm V}$
at three core radii (30 arcmin) turned out to be in the range of 350-1000. For this
distance our isotropic models give $M/L_{\rm V}=593, 369$ and $250$ for
$\alpha=0, 1$ and $3/2$ respectively.

\subsection{Anisotropic orbits}

Comparing the predictions shown in upper panels of Figures~\ref{fisomtl}
and \ref{disomtl} with the data
we find that the cuspy profiles have to be aided by a certain amount of
tangential anisotropy in order to better
reproduce the shape of the observed velocity dispersion profiles.
Trying more radial anisotropy than in the case of isotropic orbits by
using positive values of $\beta$ makes the curves decrease even more steeply
with distance, contrary to the
trend observed in the data. In the case of the profile with $\alpha=0$ it is
not clear whether more radial or more tangential anisotropy would provide a better
fit. Therefore in what follows we perform a two-parameter fitting of the velocity
dispersion profiles with velocity anisotropy $\beta=$const as a second
free parameter in addition to the dark mass, $M_v$.

The results of such least squares fitting procedure of the Fornax and Draco data
are shown in Figures~\ref{mvchbaf} and \ref{mvchbad}. The Figures are analogous to
Figures~\ref{mvchiif} and \ref{mvchiid} for the isotropic case except for the
added uppermost panels showing the best-fitting $\beta$ parameters with their
1$\sigma$ error bars. The estimated dark virial masses in the middle panels
are now roughly the same for Fornax and somewhat higher for Draco than for
isotropic orbits while their errors are now 17-50\% for Fornax and
31-105\% for Draco for the whole range of concentrations. The quality of fits
(bottom panels) in terms of $\chi^2/N$ (here $N=12$ for Fornax and $N=3$ for Draco)
is now significantly better for cuspy profiles and also somewhat
improved for the profiles with $\alpha=0$. However, the profiles with cores
produce best-fitting $\beta$ closest to zero.

Figures~\ref{fanimtl} and \ref{danimtl} are the anisotropic analogues of
Figures~\ref{fisomtl} and \ref{disomtl}. As in the isotropic case for
$\alpha=0$ profile we have chosen $c=30$ in the case of Fornax and $c=50$ in
the case of Draco, while for cuspy profiles we have taken $c=20$ for $\alpha=1$
and $c=10$ for $\alpha=3/2$. These values of concentrations happen to be those where
$\chi^2/N$ reaches minimum for different profiles in the case of Fornax (see the lowest
panel of Figure~\ref{mvchbaf}).
Upper panels of Figures~\ref{fanimtl} and \ref{danimtl} prove that
when arbitrary anisotropy parameters are allowed acceptable fits can be obtained
with all profiles. However, cuspy profiles require significantly more tangential
anisotropy, i.e. lower $\beta$ than the profile possessing a core.

The best-fitting values of $M_v$ and $\beta$ used in Figures~\ref{fanimtl}
and \ref{danimtl} are given in the lower part of Table~\ref{parameters} together
with their
errors following from the uncertainties in velocity dispersion measurements. As in
the isotropic case the errors in brackets for $M_v$ include the effect of uncertainties
in luminosity, distance and stellar mass-to-light ratio. The error estimates of
$\beta$ are not significantly affected by these additional factors and are not listed.
It is clear from Table~\ref{parameters} that in all cases
the measurements of velocity dispersion
are the dominant source of error, while other sources affect the results only slightly
and more for dark matter profiles with lower $\alpha$. One may object, however, that the
distance errors given in Table~\ref{tab1} are rather small and do not encompass 
higher distance estimates by other authors (Mateo 1998b). 
Increasing the adopted distances by 10 kpc would  decrease the virial masses given 
in Table~\ref{parameters} by up to 10\% for Fornax and 40\% for Draco for $\alpha=0$ 
with almost no difference for $\alpha=3/2$ (the estimates of $\beta$ would remain roughly
the same).

{\samepage
\begin{figure}
\begin{center}
    \leavevmode
    \epsfxsize=7.8cm
    \epsfbox[50 50 320 790]{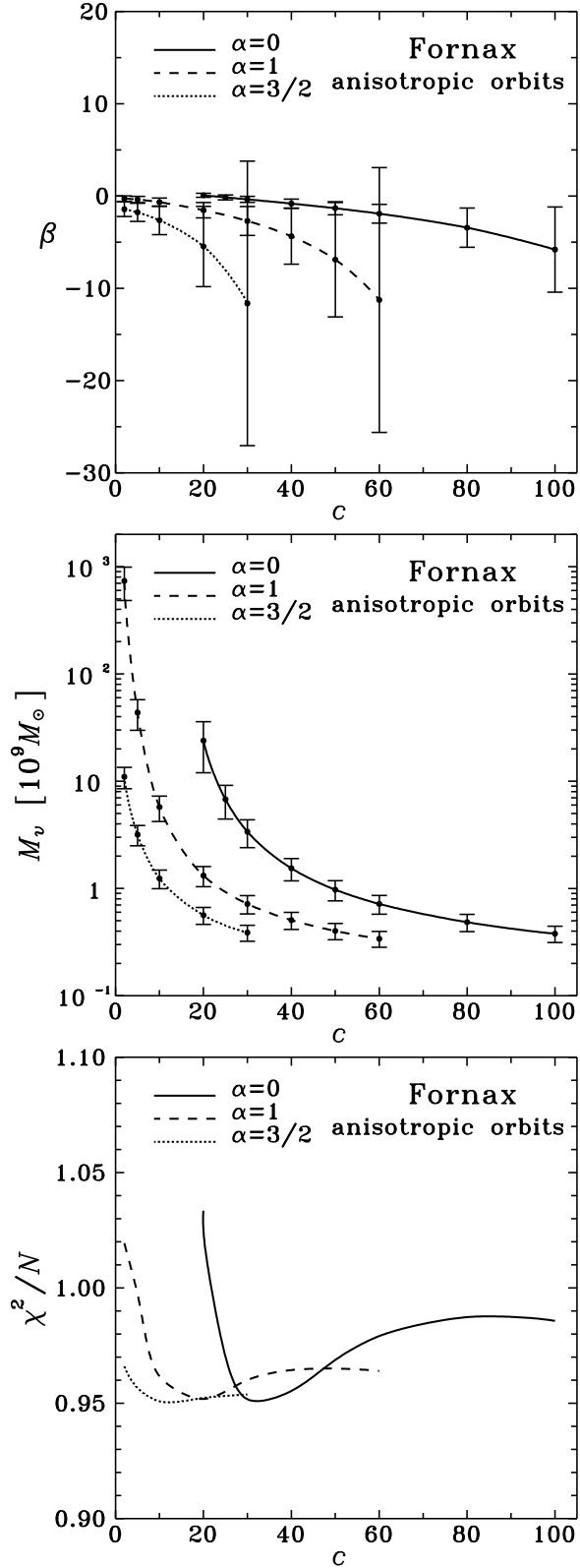}
\end{center}
\caption{The best fitting anisotropy parameters $\beta$ (upper panel)
and masses of dark haloes (middle panel) as a function of
assumed concentration parameter for profiles with different $\alpha$ in the case
of anisotropic orbits for the Fornax dwarf. Lower panel: the goodness of fit
measure, $\chi^2$, for the fits in the upper panels.}
\label{mvchbaf}
\end{figure}

\begin{figure}
\begin{center}
    \leavevmode
    \epsfxsize=7.8cm
    \epsfbox[50 50 320 790]{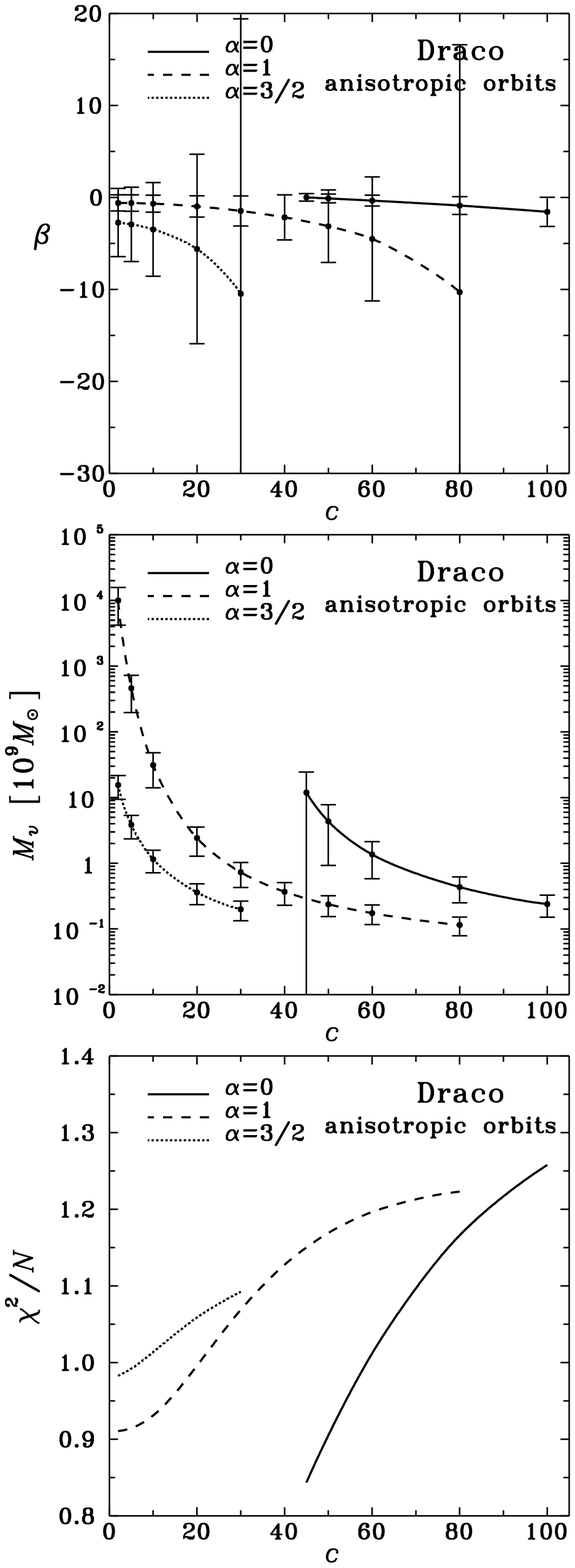}
\end{center}
\caption{Same as Figure~\ref{mvchbaf}, but for Draco.}
\label{mvchbad}
\end{figure}
}

{\samepage
\begin{figure}
\begin{center}
    \leavevmode
    \epsfxsize=7.8cm
    \epsfbox[50 50 320 560]{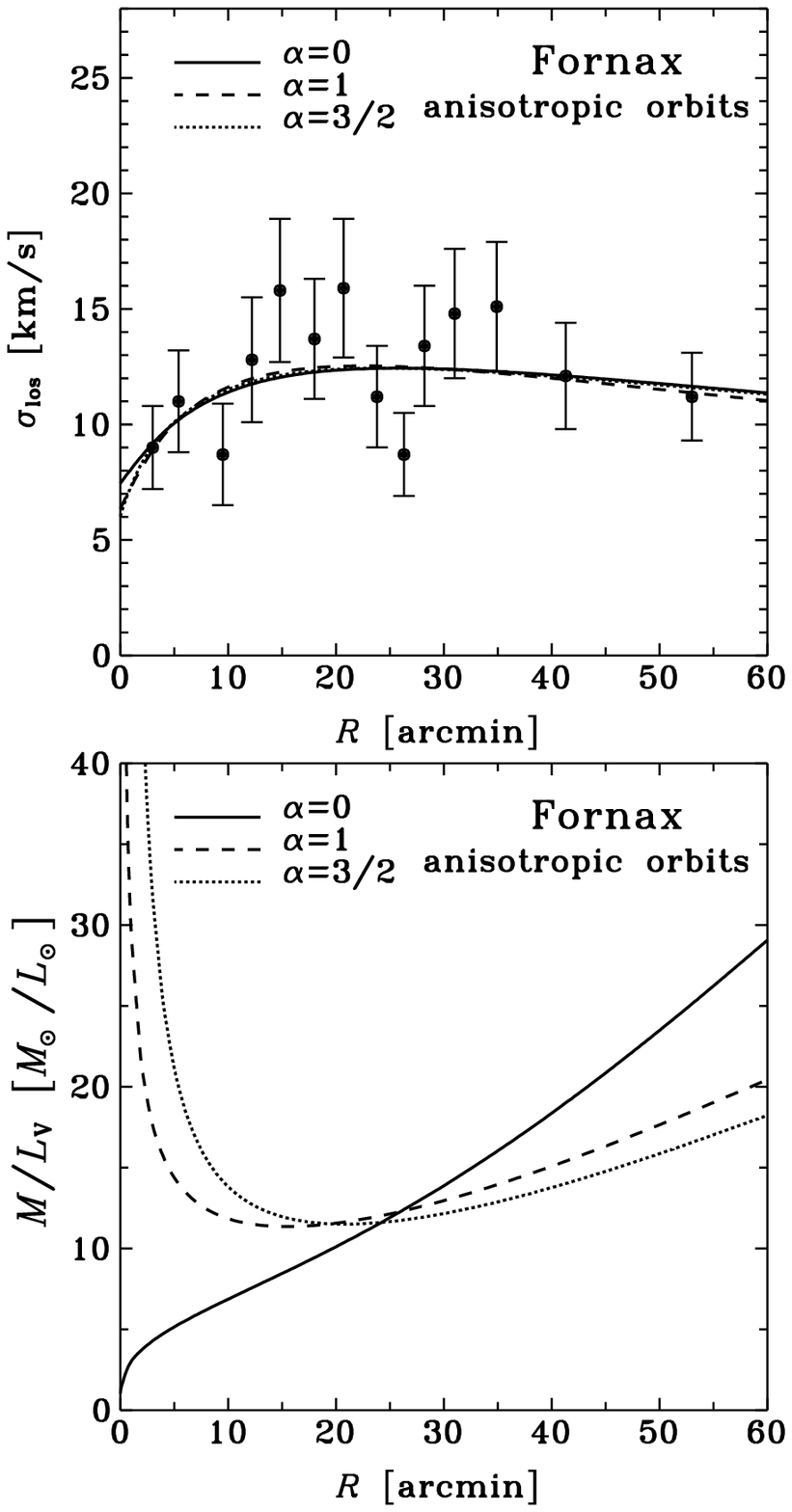}
\end{center}
\caption{Upper panel: the best fitting velocity dispersion profiles
obtained for $c=30$ ($\alpha=0$), $c=20$ ($\alpha=1$) and $c=10$ ($\alpha=3/2$)
in the case of anisotropic orbits for the Fornax dwarf. Lower panel:
mass-to-light ratios for the fits in the upper panel.}
\label{fanimtl}
\end{figure}

\begin{figure}
\begin{center}
    \leavevmode
    \epsfxsize=7.8cm
    \epsfbox[50 50 320 560]{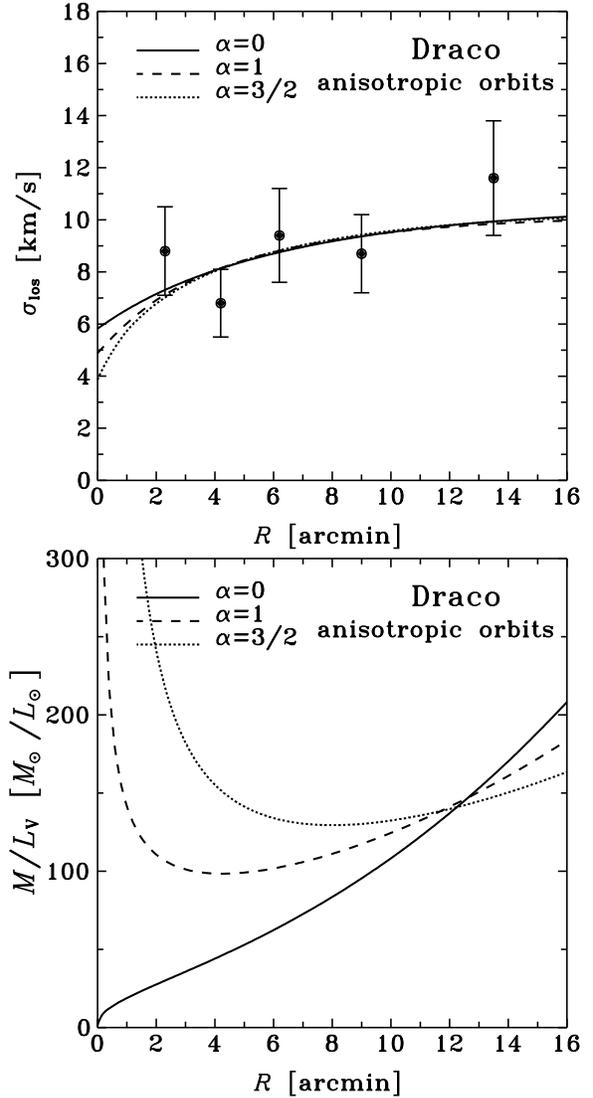}
\end{center}
\caption{Same as Figure~\ref{fanimtl}, but for Draco. The best fitting
velocity dispersion profiles plotted were obtained for $c=50$ ($\alpha=0$),
$c=20$ ($\alpha=1$) and $c=10$ ($\alpha=3/2$). }
\label{danimtl}
\end{figure}
}

One may also wonder how the best-fitting parameters are affected by our definition of
the virial mass. It turns out that decreasing the parameter $v$ in equation (\ref{d2})
by a factor of two changes mainly best-fitting $M_v$ (leaving $\beta$ roughly the same)
for the same value of concentration by increasing it typically by a factor of a few, more
for lower $\alpha$ and lower $c$ (it must be emphasized however, that $c$ given by
equation (\ref{d1}) would then be a different quantity since $r_v$ for a given $M_v$
would be increased).

The mass-to-light
ratios shown in the lower panels of Figures~\ref{fanimtl} and \ref{danimtl} do not
change significantly with respect to the isotropic case and the very different
behaviour for cuspy profiles versus the profile with a core is preserved.
In the case of Draco at $R=30$ arcmin our anisotropic models give $M/L_{\rm V}=606,
414$ and $300$ for
$\alpha=0, 1$ and $3/2$ respectively, values slightly higher than in the isotropic
case. The values agree to the order of magnitude with the estimate given by Kleyna et
al. (2002) who find mass-to-light ratio of 330 for their best-fitting anisotropic
model of Draco.

\subsection{Predictions for the kurtosis}

\begin{figure}
\begin{center}
    \leavevmode
    \epsfxsize=7.8cm
    \epsfbox[50 50 320 560]{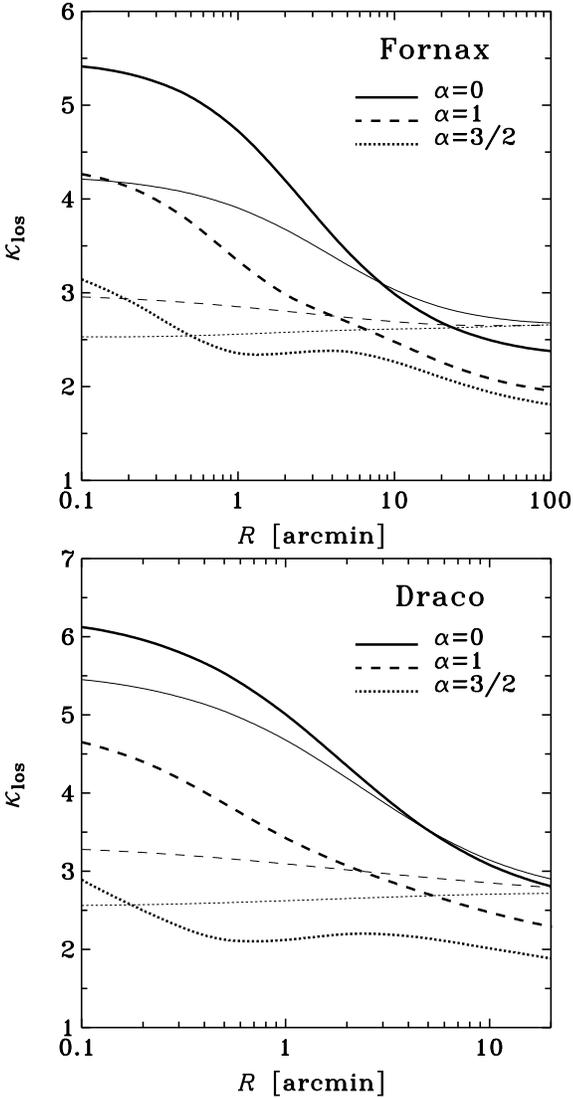}
\end{center}
\caption{Kurtosis predictions for the best fitting models of Table~\ref{parameters}
for Fornax (upper panel) and Draco (lower panel). Thick lines show the results for
anisotropic models while thin lines for isotropic ones.}
\label{k}
\end{figure}

In the previous subsection we have shown that when velocity anisotropy is treated
as a free parameter equally good fits to velocity dispersion profiles can be
obtained with both cuspy profiles and the profile with a core. In order to get rid of
this degeneracy and discriminate between the profiles one has to resort to higher
order moments of the velocity distribution such as the kurtosis introduced in
Section~2. Figure~\ref{k} shows the predictions for this quantity for models which
turned out to provide best fits to velocity dispersion profiles of Fornax and Draco.
The parameters used for every type of profile were taken from Table~\ref{parameters}.
In each panel the thick lines show the predictions for anisotropic models while thin
lines for isotropic ones.

The results prove that the kurtosis can indeed differentiate between the profiles with
a cusp and those with a core. In the anisotropic case we find for both galaxies that
the profile with a core results
in a higher value of the kurtosis at a given distance from the centre and its
value decreases more steeply with $R$ than in the case of cuspy profiles. The situation
is similar in the case of isotropic orbits, but here $\kappa_{\rm los} (R)$ definitely
decreases with distance only for the $\alpha=0$ profile while its values remain almost
constant for the cuspy profiles (slightly decreasing for $\alpha=1$ and slightly
increasing for $\alpha=3/2$).

\section{Alternative: MOND}

DSph galaxies have low densities and therefore they lie in the regime of small
accelerations. According to Modified Newtonian Dynamics (MOND), the theory
proposed by Milgrom (1983), for accelerations below the characteristic
scale of $a_0 \approx 10^{-8}$cm/s$^2$ the laws of Newtonian dynamics may
change affecting either inertia or gravity and mimic the presence of dark matter.
Although originally designed to explain flat rotation curves of spiral galaxies, MOND
is expected to operate in all low-acceleration systems. It has been e.g. shown
to explain many observed features of LSB galaxies by McGaugh \& de Blok 1998. On the
other hand, Lake (1989) claimed MOND to be inconsistent with rotation curves of a few
dwarf galaxies (the required values of $a_0$ being different for different systems
and generally too low).
In the case of dSph galaxies MOND should cause an increase of velocity dispersions.

\begin{table}
\caption{Best-fitting parameters for MOND in the isotropic and
anisotropic case with 1$\sigma$ error bars. $a_0$ is given in units
of $10^{-8}$cm/s$^2$, while $M/L_{\rm V}$ in $M_{\sun}/L_{\sun}$.}
\label{mond}
\begin{center}
\begin{tabular}{ccccc}
case & assumed & fitted & Fornax & Draco \\
\hline
$\beta=0$ & $M/L_{\rm V}=1$ & $a_0$         & $2.2 \pm 0.5 (0.8)$ & $39 \pm 13 (22) $ \\
          &  $a_0=1.2$        & $M/L_{\rm V}$ & $1.8 \pm 0.4 (0.7)$ & $32 \pm 10 (18) $ \\
\hline
$\beta \ne 0$ & $M/L_{\rm V}=1$ & $a_0$   & $2.1 \pm 0.5 (0.8)$  & $50 \pm 19 (29)$ \\
              &                 & $\beta$ & $-0.96 \pm 0.53$ & $-1.5 \pm 1.6$ \\
              & $a_0=1.2$       & $M/L_{\rm V}$ & $1.8 \pm 0.4 (0.7)$ & $40 \pm 14 (23)$ \\
              &                 & $\beta$ & $-0.97 \pm 0.53$ & $-1.5 \pm 1.7$ \\
\hline
\end{tabular}
\end{center}
\end{table}

It is presently far from clear, whether MOND can explain the observed velocity
dispersions of dSph galaxies. Although for most dwarfs it gives satisfactory results,
notorious problems have been caused by Draco. Analyzing its central velocity
dispersion Gerhard \& Spergel (1992) have found it incompatible with stellar
mass-to-light ratios, contrary to the conclusion of Milgrom (1995) who claimed
that its mass-to-light ratio is consistent with stellar values
if all the observational errors are properly taken into account.

In the previous work (\L okas 2001) we have shown how to generalize the method
of predicting the line-of-sight velocity dispersion profile to MOND and applied it
to fit the data for Fornax, Draco and Ursa Minor dwarfs. We have assumed the
stellar mass-to-light ratio of $M/L_{\rm V}=1 M_{\sun}/L_{\sun}$ and fitted parameters
$a_0$ and $\beta$. The best-fitting values of $a_0$ were different for different
dwarfs, acceptable for Fornax, but much higher than expected in the case of Draco and
Ursa Minor. We concluded that while the case of Ursa is doubtful since
 it has been recently observed to possess tidal tails (Mart\'{\i}nez-Delgado
et al. 2001), the high value of $a_0$ estimated for Draco can pose a
serious problem for MOND (a similar conclusion based on the analysis of a velocity
dispersion profile of Draco from a different data set was also recently reached by
Kleyna et al. 2001).

We have repeated the analysis here for Fornax and Draco with the least-square
fitting method, similar to
that applied to fit dark matter profiles, in order to provide the error estimates for
the fitted parameters. Both isotropic and anisotropic cases were studied. The
results of the fitting procedure are shown in Table~\ref{mond} with the isotropic case
in the upper part and the anisotropic in the lower. We assumed
$M/L_{\rm V}=1 M_{\sun}/L_{\sun}$ and fitted the acceleration parameter $a_0$ in
one case while taking $a_0 = 1.2 \times 10^{-8}$cm/s$^2$, the value known to fit
best the rotation curves of spiral galaxies (Begeman, Broeils \& Sanders 1991;
Sanders \& Verheijen 1998), and finding $M/L_{\rm V}$ in the other.

As is clear form Table~\ref{mond} both approaches yield acceptable values of
$M/L_{\rm V}$ or $a_0$ in the case of Fornax, while very high numbers in the case of
Draco. However, the errors for Draco estimated from the uncertainties in the velocity
dispersion measurements are so large that the acceptable values of $M/L_{\rm V}$
or $a_0$ (of the order of unity in the units of the Table) are still within
3$\sigma$ error bars. Analysis of the velocity dispersion profile therefore
reduces the error
significantly compared to what can be obtained from a single central velocity
dispersion measurement (then acceptable $M/L_{\rm V}$ are within 2$\sigma$ error from
the best-fitting value, McGaugh, private communication), but still not enough to
rule out MOND. In fact
these errors are farther increased if the errors in luminosity measurement
(quite big in the case of Draco, see Table~\ref{tab1}) and in distance
are taken into account. The resulting larger errors (due mainly to the error
in luminosity) are given in brackets in Table~\ref{mond}. With these
errors the values obtained here are safely within 2$\sigma$ of the acceptable ones.
(A discussed in Section~4 this source of error is not important for dark matter
because it dominates strongly the distribution of stars.)
We therefore conclude that with present quality of the data the alternative
explanation of velocity dispersions in terms of MOND cannot yet be ruled out.

\section{Discussion}

We presented predictions for the line-of-sight velocity dispersion profiles
of dwarf spheroidal galaxies and compared them to observations in the case
of the Fornax and Draco dwarfs. We discussed different dark matter distributions,
both cuspy and possessing flat density cores. For isotropic models the
dark haloes with cores are found to generally fit the data better than those
with cusps. The fits for each type of profile tend to be better for lower
concentrations. For cuspy profiles those cannot be taken arbitrarily low
since, as discussed in Subsection 3.2, $N$-body simulations (providing the
basis for the choice of profiles here) predict higher concentrations for smaller
masses. Besides, fits with low concentration require incredibly high
corresponding dark masses of the order of a normal galaxy. In our exemplary
cases we have taken $c=10$ for $\alpha=3/2$ and $c=20$ for $\alpha=1$.
The dark matter profiles with such parameters yield for Draco fits
with $\chi^2/N=1.6$ and $\chi^2/N=1$, respectively, while an $\alpha=0$ profile
with $c=50$ has $\chi^2/N=0.6$. The situation is similar for Fornax.

Anisotropic models were studied by fitting
two parameters, dark mass and velocity anisotropy, to the data.
We find that the steeper the cusp of the profile, the more tangential is the
velocity distribution required to fit the data, in agreement with the
well-known degeneracy of density profile versus velocity anisotropy.
In this case acceptable fits (with $\chi^2/N=1$ or lower)
can be obtained for all considered profiles, but
only the profiles with cores yield good fits with anisotropy
parameter close to the isotropic value. To discriminate between different profiles
we offer predictions for the kurtosis of the line-of-sight velocity distribution.
Our findings in the case of the velocity dispersion profile for Draco
are complementary to and in qualitative agreement with those of Kleyna et al.
(2002) who tested dark matter distributions varying at large distances. Their
best-fitting models required the halo to be nearly isothermal (as our models are
in a wide range of distances) and the velocity anisotropy to be weakly tangential.

The amount of dark
matter inferred depends on the concentration parameter of the density profile, but
for realistic concentrations is typically of the order of $10^9 M_{\sun}$ both
for Fornax and Draco for all profiles and velocity distributions.
Since the total luminosity of Draco is two orders of magnitude lower than that
of Fornax (see Table~\ref{tab1}), this results in a significant
difference in their corresponding mass-to-light ratios. At the scale of the order of
two S\'ersic radii Fornax has $M/L_{\rm V} \approx 10 M_{\sun}/L_{\sun}$ while for
Draco we obtain $M/L_{\rm V} \approx 100 M_{\sun}/L_{\sun}$. However, mass-to-light
ratio dependence on distance is very different for different profiles. For
cuspy profiles $M/L_{\rm V}$ diverges both at small and at large distances from the
centre, while for $\alpha=0$ profile it is finite at the centre and grows steadily
with distance. For distances larger than two S\'ersic radii profiles with
lower $\alpha$ give higher values of $M/L_{\rm V}$ and the anisotropic models yield
slightly higher $M/L_{\rm V}$ than isotropic ones.
As discussed in Section 3, our results for $M/L_{\rm V}$ outside the
galactic core agree to the order of magnitude with the estimates of Kleyna et al.
(2001, 2002).

A possible source of uncertainty in the analysis of velocity dispersions is the
influence of binary population. In the recent study De Rijcke \& Dejonghe (2002)
using realistic distributions of orbital parameters estimated the maximum additional
velocity dispersion due to binary stars to be of the order of $3$ km/s (in
agreement with earlier studies). It turns out
that only stellar systems with intrinsic dispersions close to this value (e.g.
globular clusters) will be affected. For velocity dispersions characteristic of
dSph galaxies studied here the effect is small: the velocity dispersion $\sigma=9$
km/s for a galaxy with a Gaussian line-of-sight velocity distribution can increase
due to the presence of binaries by at most 5\% (for the maximum binary fraction of
$1$, see Figure~15 of De Rijcke \& Dejonghe 2002), much less than the typical
error in velocity dispersion measurements. The binary fraction of Draco is
estimated to be between 0.2 (Olszewski, Pryor \& Armandroff 1996)
and 0.4 (Kleyna et al. 2002), which would
make the effect even smaller. For Fornax no information on binaries is yet
available, but the typical velocity dispersions are higher in this galaxy which
makes the effect less important. We therefore expect binaries to only slightly
affect our best-fitting models so as to increase the estimated virial mass. More
influence is expected in the case of kurtosis: its Gaussian value
$\kappa_{\rm los}=3$ for the same dispersion $\sigma=9$ km/s will increase by up to 23\%.
Hence when using the kurtosis predictions of Subsection~4.3 to discriminate between
different dark matter profiles the influence of binary population will have to be
studied in more detail.

The analysis presented in this paper is valid under the condition that dSph
galaxies are in dynamical equilibrium. However, those of them studied here,
as members of the Local Group exist in gravitational field of larger galaxy, the
Milky Way. One may worry that the tidal interactions with bigger galaxy may affect the
dynamics of some dwarfs and our interpretations of their velocity
dispersions. Some of the dwarfs have been indeed observed to possess tidal tails, e.g.
Ursa Minor (Mart\'{\i}nez-Delgado et al. 2001), Sagittarius (Ibata, Gilmore \&
Irwin 1994) or Carina (Majewski et al. 2000). However, in the case of Draco dwarf
studied here no evidence of tidal interactions was found (Piatek et al. 2002).

Tidal interactions have been studied
theoretically by numerical modelling with conflicting results.
Piatek \& Pryor (1995) have performed numerical simulations of such effects and
concluded that they should not affect much the inferred mass-to-light ratios.
However, Kroupa (1997) and Klessen \& Kroupa (1998) have demonstrated
that what appears to
be a dSph galaxy in equilibrium may in fact be  a tidal remnant displaying high
velocity dispersions without any dark matter. As suggested by Oh, Lin \& Aarseth
(1995) it may also very well be that some dwarfs indeed contain dark matter while
others may be part of tidal debris. It is still debated, whether tidal interactions
are common among dSph galaxies and whether they
may cause an increase of velocity dispersion that is likely to be interpreted as
presence of dark matter.

\section*{Acknowledgements}

I wish to thank P.
Salucci for encouragement to follow the line of investigation presented in this
paper and S. McGaugh for discussions on dSph galaxies in MOND. Extensive comments
from an anonymous referee helped to significantly improve the paper. This research was
partially supported by the Polish State Committee for Scientific Research
grant No. 2P03D02319.

\end{document}